\documentclass[12pt]{article}

\usepackage{epsfig,graphicx}
\newcommand{\be}{\begin{equation}}
\newcommand{\ee}{\end{equation}}
\newcommand{\bea}{\begin{eqnarray}}
\newcommand{\eea}{\end{eqnarray}}

\begin{document}
\newcommand{\notE}{E\kern-0.6em\hbox{/}\kern0.05em}
\begin{titlepage}
\begin{flushright}
MCTP-05-72\\
\end{flushright}
\vspace{0.5cm}
\begin{center}
{\Large \bf Outside the mSUGRA Box}\\
\vspace{1cm} \renewcommand{\thefootnote}{\fnsymbol{footnote}}
{\large Jacob L. Bourjaily$^*$\footnote[3]{Email:
jbourj@umich.edu},
 Gordon L. Kane$^*$\footnote[5]{Email: gkane@umich.edu},
 Piyush Kumar$^*$\footnote[2]{Email: kpiyush@umich.edu},
 Ting T. Wang$^*$\footnote[4]{Email: tingwang@umich.edu}} \\
\vspace{1.cm} \renewcommand{\thefootnote}{\arabic{footnote}} {\it
* Michigan Center for Theoretical Physics (MCTP)\\ Ann
  Arbor, MI 48109, USA \\
}
\end{center}
\vspace{0.5cm}


\begin{abstract}
Most studies of the potential for discovery of superpartners at the Fermilab
Tevatron or CERN LHC have focused on the so-called mSUGRA (minimal supergravity mediated
supersymmetry breaking) model, not because
it is well motivated but because it has a minimal number of parameters. If
signals are seen that could be superpartners, most analyses will
attempt to interpret them in the mSUGRA framework since the needed software 
and computational tools exist. With only a few signal channels, and initially large statistical and
systematic errors, it is very likely that an mSUGRA interpretation will look
all right even if it is not. We present an approach to studying any potential
signals of new physics in ``inclusive signature space'' that
sensitively tests any proposed interpretation, and apply it to the mSUGRA
case. The approach also has significant experimental advantages, reducing
the sensitivity to jet energy corrections, dependence on beam luminosity, 
and other systematics. Basically, if one (or more) instances of
reported data lies outside certain bounded regions of inclusive signature space 
characteristic of the
physics being tested, that physics is excluded for any parameters. The
approach can be used to study supersymmetry breaking and to point to the
form of the underlying theory even without detailed measurements of a number of 
important parameters which will be difficult or impossible to measure at hadron colliders.
\end{abstract}
\vspace*{1cm}
\end{titlepage}

\section{Introduction}
It is exciting that we are getting closer to the time when direct signals of
new particles or interactions at hadron colliders may give us data that will
start to point toward how the foundations of the Standard Model are
strengthened, and how the SM is extended. Such information could come from
the Fermilab Tevatron collider as it accumulates data. If the indirect
clues we have today are not very misleading such effects are very likely to
come at the LHC. For simplicity in this paper we will focus on the LHC.

We will assume that the new physics is supersymmetry at
the weak scale,
which is well motivated. However, the approach we
advocate would also be relevant if some other new physics were discovered.
The experimental discovery of supersymmetry would be manifest by signals
interpretable as the superpartners of the SM particles. Since supersymmetry
is a broken symmetry, the masses of the superpartners are not known. But
otherwise the full Lagrangian for the MSSM is known, and for a given set of masses all
superpartner production and decays can be calculated. For this paper we
also assume the lightest superpartner (LSP) is stable and escapes the
detector; a similar analysis could be done if the LSP is unstable. One of
the main goals once there is data will be learning how supersymmetry is
broken. Even if that is not known, the theory is very constraining, and it
is known how to write the general form of the Lagrangian. Ideally data
would allow us to measure the terms of the so-called general ``soft-breaking''
Lagrangian. But if we only have data from hadron colliders, the number of
observables is always less than the number of mass parameters in the Lagrangian, so in
general data cannot be inverted to deduce the Lagrangian parameters. It is
dangerous to ``solve'' this problem by making assumptions that reduce the
number of parameters since any wrong assumption could result in very
misleading implications. One important issue is that most of the
Lagrangian masses are complex, leading not only to possibly large CP violating effects 
at colliders, but also shifting the values of superpartner masses and
branching ratios. So, if the Lagrangian masses are assumed to be real the
results could be unrealistic.

In order to learn how to interpret signals it is important to pay attention
to what experimenters actually measure. These are first cross sections
times branching ratios for production and decay of superpartners, and 
some kinematical distributions. That is all that can be measured at
colliders. The soft-breaking Lagrangian on the other hand depends on
``soft masses''. The soft masses in general are not physical mass eigenstate
masses, which usually result from diagonalizing mass matrices that depend on
the soft masses in complicated ways. A top-down approach would go from an
underlying theory to a superpotential and a K\"{a}hler potential and a gauge
kinetic function, from which the soft-breaking Lagrangian can be calculated.
Then the mass matrices for the various sectors of the theory would be
written, and the mass eigenstate masses and their production and decay rates
calculated. These would imply certain experimental distributions. The
mass eigenstates themselves are only measurable in certain special cases for
a few eigenstates. Thus in general experimenters will present a dozen or so
rates, and a few kinematical distributions. How can one interpret such
information to deduce the Lagrangian and study its patterns and properties
in order to learn about how supersymmetry is broken and about the underlying
theory?

If we had a large amount of precise data, particularly from a linear
collider, in principle one could take a bottom up approach from the data to
the theory. But for at least a decade after LHC begins to take data, and perhaps much
longer, we will only have hadron collder data, which does not generally permit
going from observables to the full Lagrangian. For the first few years
the signals will have significant statistical and systematic errors. Under
those conditions one will be able to describe the data with a variety of
models, each of which has a few parameters. In practice almost all studies
so far have been done in the mSUGRA framework
\cite{Arnowitt:1992aq,Kane:1993td}, not because it is
well motivated but because it has only a few parameters. 
Because historically it was a sensible approach to study early-on, it is by
far the best studied approach. In practice, it is likely that some pieces of initial data will
be describable by an mSUGRA model even if mSUGRA is not the actual theory.

In this paper we want to advocate a sensitive and systematic approach to
interpreting a limited amount of data. We will see that the method is
potentially very powerful in answering questions that are difficult or
impossible with the usual methods. Here we will focus on the issue of
deciding whether mSUGRA can be a valid extension of the SM when new data is
available.

While mSUGRA is indeed a simple \emph{parametrization} of the soft Lagrangian,
it is not truly a
\emph{model} because it does not contain an explanation for its
various assumptions. mSUGRA corresponds to a situation in which
minimal kinetic terms are assumed for the K\"{a}hler potential in the 
supergravity lagrangian. It turns out
that these assumptions are hard to realize in a fundamental
microscopic theory like string theory. Therefore, from a
theoretical point of view, mSUGRA should be extended. For example, the origin of the 
auxiliary field
vacuum expectation values and the assumed reality of the soft terms should
be explained, the neutrino sector should be added, and the
smallness of $\mu$ should be explained dynamically, just to
mention a few things.

On the other hand, from the phenomenological point of view, mSUGRA
is sometimes considered to be attractive due to its simplicity and 
most phenomenological studies of collider as well as non-collider
(flavor physics, CP physics, cosmology) observables are based on
the framework of mSUGRA. However, even in this special case, although a
lot of work has been done to learn how to measure properties of
new physics at the LHC, most of it relies on implicit assumptions.
There are many obstacles to determining the relevant properties
in a general case. Let us try to analyze some of these obstacles
closely.

\section{Obstacles}

One major obstacle is that experiments measure cross-section
times branching ratios ($\sigma \times BR$), and masses of mass
eigenstates (mostly only mass differences). None of these
quantities appear in the Lagrangian. The chargino, neutralino and
stop masses for example, are related to the complex soft
parameters $M_1,M_2,\mu$ and the stop left and right handed mass
parameters in an indirect way, by being the eigenvalues of the
appropriate mass matrix. Even though the gluino doesn't mix with
others in a mass matrix, it gets appreciable radiative
SQCD corrections. Therefore, it is not straightforward to learn the
implications of these measured quantities.

A second obstacle is that at a hadron collider there are always
more Lagrangian parameters than observables. Hence in general it is
not possible to solve for all Lagrangian parameters such as
$\tan\beta,\mu,$ gaugino masses, scalar masses, trilinear
couplings, etc. This general problem has been largely overlooked
and was first studied in \cite{Chung:2003fi,Kane:2002tr}. The usual
approach to this problem is to keep making {\it ad hoc} assumptions
until there are fewer parameters than observables to ``solve'' for
the parameters. However, unless those assumptions are correct, this
approach may not give the right answers.

A third obstacle arises due to techniques usually employed to
analyze data. In order to measure masses and production
cross-sections times branching ratios, selections and
cuts are imposed on data to reduce backgrounds and isolate
signals. In many cases, the statistics is limited and this may 
reduce the signal so much that little can be learned.
Also, the available channels are mostly analyzed separately, one
at a time. In such a situation, issues about absolute
measurements, beam luminosity, etc. have to be taken into account.
For example, jet energy corrections that affect missing transverse
energy become quite important. In addition, at the LHC, the large
number of channels may make separation of states very difficult.

Taking mSUGRA as an example, suppose a signal is found at the LHC
in some channel, using cuts imposed on data. A traditional
``single channel analysis'' would mean that the statistics are very
limited and there are large errors due to both theoretical and
experimental uncertainties. In such a situation, it would most
likely be possible to find a point from the mSUGRA parameter space
which fits the data (with large uncertainties) even if it were not the
correct theory. 

Therefore, a new innovative approach is needed to overcome these
obstacles and to devise ways to favor a particular class of models
while weeding out most of the others, once there is data. The
philosophy of one such approach was spelled out in
\cite{Binetruy:2003cy}. We implement it in detail in the following
sections.

\section{A Constructive Approach}

The approach is two pronged. The first is consistency. If an
approach to Beyond-the-Standard Model physics (like mSUGRA) producing a 
particular set of
masses can reproduce a particular signal, it implies a variety
of other constraints since different processes involving the same
masses have to be consistent with each other. The more data one
adds from other processes, the more one constrains the parameters.
Therefore, a ``multi-channel analysis'' is called for.

The second is to use information obtained from the pattern of
\emph{inclusive} signatures in various channels to learn more
about the theory. By an `inclusive signature', we mean any signature
observable in experiments indicating the existence of physics
beyond the standard model, summed over all possible ways that such
a signature may arise. The basic idea is that even though a single
signature will not tell us much about supersymmetry breaking or
the underlying theory, patterns of several inclusive signatures
will. The list of inclusive signatures \cite{Binetruy:2003cy} 
should be expanded to get
more and more information about the theory.

Our approach also has implications for experimental issues. By
comparing inclusive signatures, issues that affect individual
absolute measurements, such as beam luminosity, jet energy
corrections, etc. drop out or become less important if one is only
comparing collider event rates. It may not
even be necessary to determine actual cross-sections.
Therefore, it is possible to add all the ways to get a given
signature, with very few cuts, to get large statistics. The Standard Model (SM)
contributions to the signal can and probably should be included in the comparison prediction.

\section{Application to mSUGRA}

In this section, we illustrate our approach by applying it to
mSUGRA --- the most popular framework for
phenomenological studies. The procedure is the following:

First, a sufficient number of models are generated so
as to approximately cover the parameter space of mSUGRA. Of these, only some 
will be consistent with all the current bounds for
superpartner masses, the higgs mass, constraints from flavor and
CP physics, as well as the latest relic density upper bound from WMAP.
Considering only those models consistent with current constraints, we then
compute the spectrum of
particles at the electroweak scale and inclusive signatures for each. 
These collider inclusive signatures denote number of
events expected in excess of the standard model. All the signatures used
are real observables. With inclusive signatures computed for thousands of models, we 
then make scatter plots between
different combinations of inclusive signatures. Since the
parameter space of mSUGRA is a \textit{bounded} region, it is
expected that scatter plots of different combinations
of inclusive signatures also occupy a \textit{bounded}
region in the ``space of inclusive signatures''. This
is confirmed by our procedure, as our plots will explicitly
show. Therefore, there is a precise sense in which the complete
set of bounded regions obtained for different combinations
of inclusive signatures can be thought of as the ``footprint'' of
mSUGRA.

This analysis has the potential to rule out mSUGRA almost immediately with limited data. 
In the event
of a discovery, signals will appear as isolated points in
the relevant inclusive signature plots. If even a \emph{single}
point lies well outside the mSUGRA region in the inclusive
signature plots, then mSUGRA has to be ruled out. On the other
hand, even if {all} the points lie within the mSUGRA region, it
does \emph{not} imply that mSUGRA is confirmed. These
issues will be explained in better detail in the following
sections.

For the present paper, for simplicity, we do not include the SM contributions to the
inclusive signatures. They are small since the transverse missing energy
$(\notE_T)$ cuts are designed to minimize them. For the full analysis in future 
that deals with real data, the SM contributions should of course be included, and the
simulations run through the detectors too. A complete study can only be done with
the actual parameters and cuts used to measure the inclusive signatures.

\section{Procedural Details}

We first generated a sufficient number of points so as
to cover the parameter space of mSUGRA, not already excluded by experiment. An mSUGRA ``point'' is
defined by five parameters at the high scale --- $m_0, \,m_{1/2},\, A_0,\,
\tan{\beta}$ and $\mathrm{sign}\,{\mu}$. The high scale was chosen to
be the unification scale $\sim 2 \times 10^{16}$ GeV. Since $\mu <
0$ is disfavored for mSUGRA \cite{Battaglia:2001zp}, only $\mu > 0$ was considered.

A set of about ten thousand mSUGRA models was used to map out the region
occupied by mSUGRA in inclusive signature space and estimate its
boundary. These were found by randomly sampling mSUGRA parameter
space for models consistent with experimental and cosmological
constraints using the framework of the most recent version of the
DarkSUSY code \cite{Gondolo:2004sc}. Each model is consistent with
constraints on $b\to s\gamma$, $(g_{\mu}-2)$, and particle
spectra. Importantly, it was also required that the thermally
produced component of the LSP relic density be no more than that
of all the dark matter in the universe, $\Omega_{LSP}h^2\leq0.15$ \cite{Spergel:2003cb}.
Because the cosmological evidence for the relic abundance of cold
dark matter is largely insensitive to its composition and the
mechanism by which dark matter was produced in the universe, there
is no lower bound on the thermal relic density of the LSP; and using a lower
bound could be very misleading.
It must be kept in mind that even in cases where the
thermal relic density of the LSP is low, it could still provide a
good dark matter candidate if mechanisms for non-thermal
production of the LSP are operative. In recent years, many non-thermal mechanisms
have been noticed in literature.

The parameters of the viable mSUGRA models obtained at the
high scale were used as an input to the Renormalization Group
(RG) evolution and superpartner spectrum calculating program ---
SUSPECT2 \cite{Djouadi:2002ze} \footnote{Only points which were
compatible with both programs --- ISAJET \cite{Paige:2003mg}
(used in DARKSUSY) and
SUSPECT2 --- were used for collider simulation.}. The collider
simulation for LHC was done by calling the Monte-Carlo simulation
program PYTHIA \cite{Sjostrand:2003wg}. A luminosity of
$10\;fb^{-1}$ was assumed for the analysis. Standard cuts from
\cite{Baer:1995va,Baer:1995nq} were applied to the output 
obtained from PYTHIA to minimize the Standard Model (SM)
background. Thus, a list of the number of events in excess of the
standard model, which passed the imposed cuts, for all relevant
collider signatures was obtained. The precise nature of cuts and
the list of signatures used is described in the following
subsection.

\subsection{Cuts}

For cuts, two basic entities must be defined: jets and leptons.
A region $|\eta|\le 3$ was picked using fixed cones of size
$R=\sqrt{\Delta\eta^2+\Delta\phi^2}=0.7$, where $\eta$ is the
pseudorapidity and $\phi$ is the azimuthal angle. Clusters in this
region with $E_T\ge 100$ GeV were labelled as jets. Leptons were
defined as electrons or muons with $|\eta|<5$ and $p_T\ge 20$ GeV.
We also required leptons to be isolated from jets by imposing that
the visible activity within a cone of $R=0.3$ about the lepton
direction is less than $E_T(\mbox{cone})=5$ GeV. These definitions
are similar to that used by Baer et al in \cite{Baer:1995va,Baer:1995nq}.

Once jets and letpons are defined, a generic signature can be
characterized as  $(m$-jet$)+(n$-lepton$)+\notE_T$, where $m$
takes 0, 1, 2, 3, and $\ge 4$ and $n$ takes 0, 1, 2 and $\ge 3$.
The transverse missing energy was required to satisfy $\notE_T\ge
100$ GeV. We also require the angle between jets and $\notE_T$
in the transverse plane to be greater than $15^{\circ}$. 
For $n=2$, the signature can be further classified as an
opposite-sign (OS) dilepton or a same-sign (SS) dilepton
signature, depending on the relative charges of the two leptons.
Thus different combinations of jet and lepton configurations give
$20$ $(=5\times 4)$ different signatures in total.

However, as already emphasized in Section 3, it is good
to use very \emph{inclusive} signatures. Here, by inclusive
signatures we mean that for a given leptonic configuration,
different jet configurations are summed over.  For example, the
``inclusive same-sign dilepton with $m\ge 2$ jets'' signature
refers to events with two same sign leptons and number of jets
greater or equal to two. The inclusive signatures we include in this
study are listed below:

\begin{enumerate}
\item 0 leptons + $\geq$ 2 jets + $\notE_{T}$ \item 1 lepton +
$\geq$ 2 jets + $\notE_{T}$ \item Opposite Sign (OS) dileptons +
$\geq$ 2 jets + $\notE_{T}$ \item Same Sign (SS) dileptons +
$\geq$ 2 jets + $\notE_{T}$ \item Trileptons + $\geq$ 2 jets +
$\notE_{T}$
\end{enumerate}

\noindent Inclusive signatures with $\geq$ 2 jets were used in our
analysis, since SM backgrounds are well known for such cases
\cite{Baer:1995va,Baer:1995nq}. In some cases, certain signatures with definite number
of jets can also be useful to probe physics. Additional signatures can be added when the analysis is done for real data.
The most important point is to only use fully observable
quantities.

\section{Plots and Analysis}

A few important plots are shown in the next few pages. The remaining plots can be seen at the website : \textbf{http://feynman.physics.lsa.umich.edu/$\sim$signaturespace}.

\begin{figure}
\center \epsfig{file=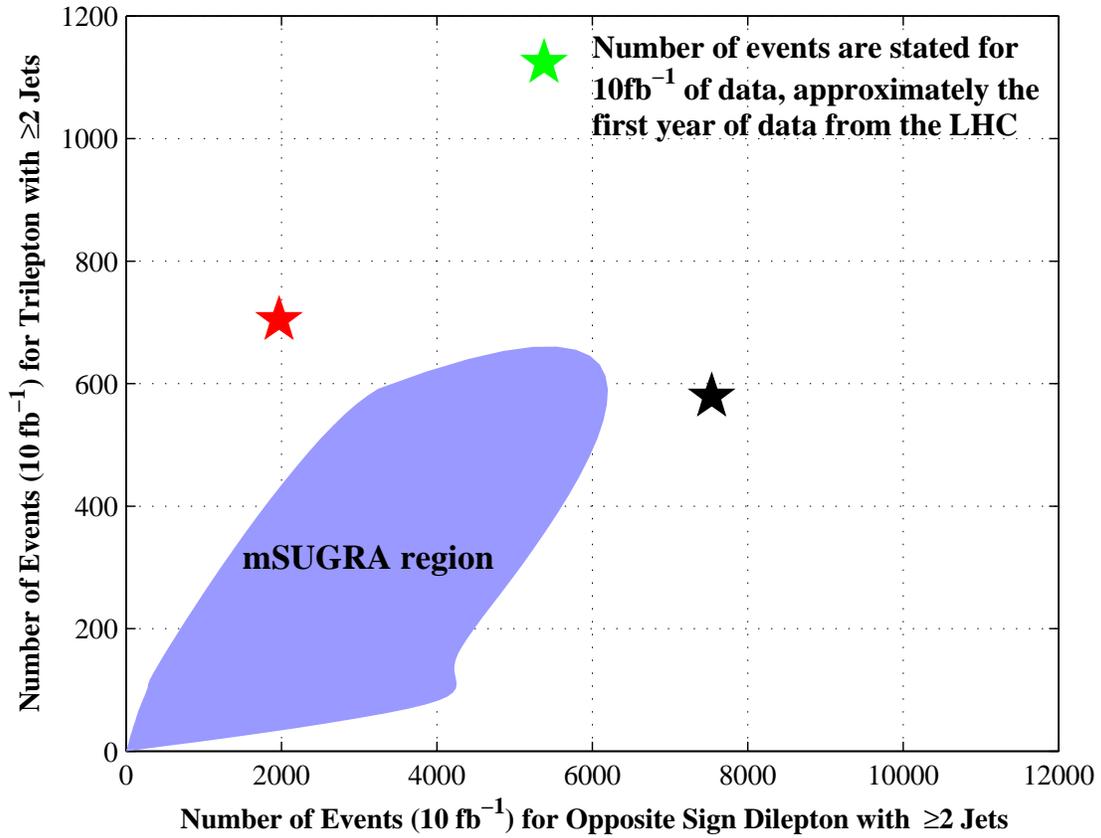,height=12cm, angle=0}
\caption{Number of events for Trilepton (3L) with $\geq$ 2
jets vs Number of events for Opposite Sign dilepton (OS) with $\geq$ 2
jets. The blue area denotes the 
region occupied by mSUGRA, while the red, green and black ``stars''
correspond to other supersymmetry breaking 
models.} \label{msugra:fig:SSvs0L}
\end{figure}

\begin{figure}
\center \epsfig{file=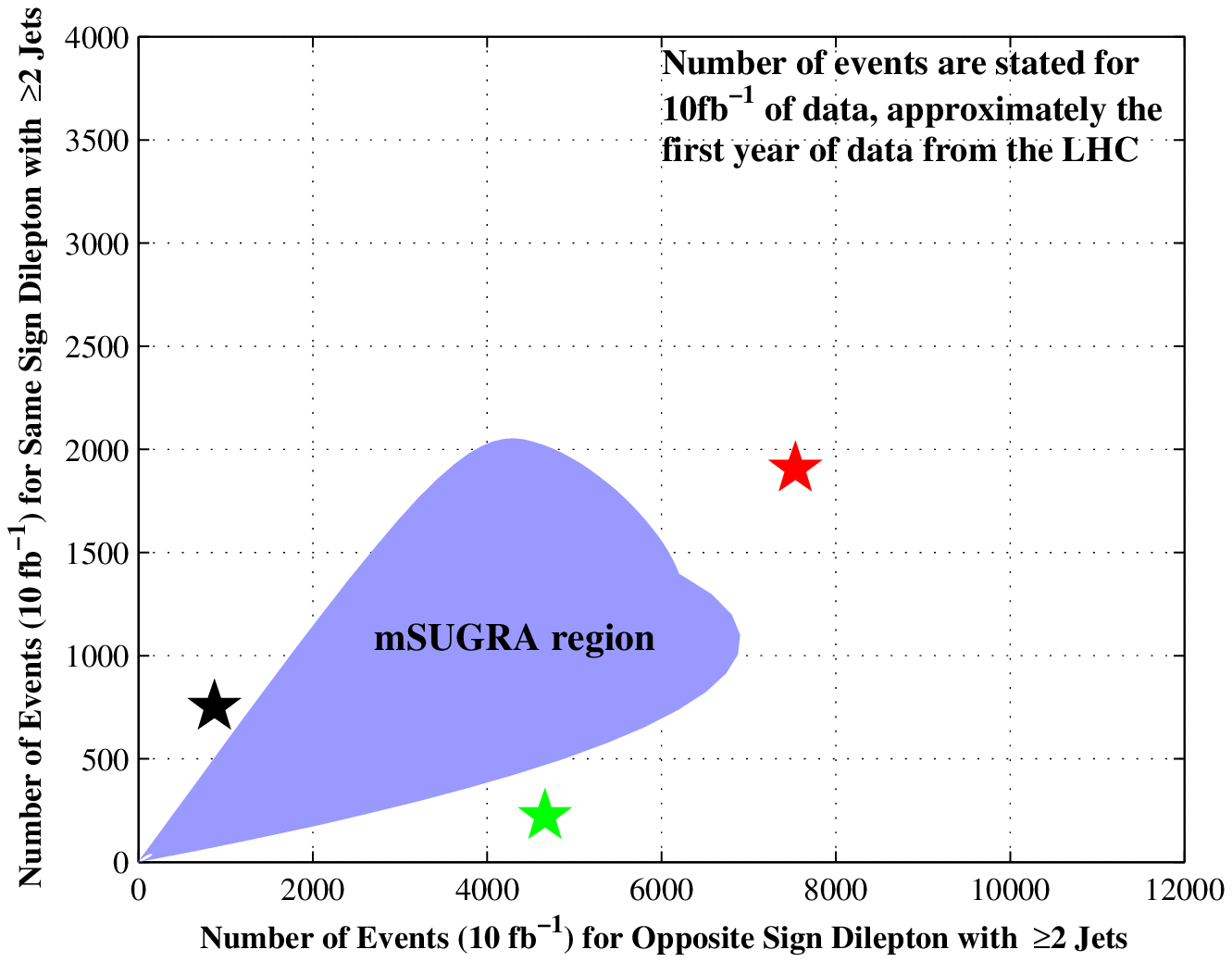,height=12cm, angle=0}
\caption{Number of events for Same Sign (SS)
Dilepton with $\geq$ 2 
jets vs Number of events for Opposite Sign Dilepton (0S) with $\geq$ 2
jets. The blue area denotes the 
region occupied by mSUGRA, while the red, green and black ``stars''
correspond to other supersymmetry breaking 
models.} \label{msugra:fig:OSvs0L}
\end{figure}

\begin{figure}
\center \epsfig{file=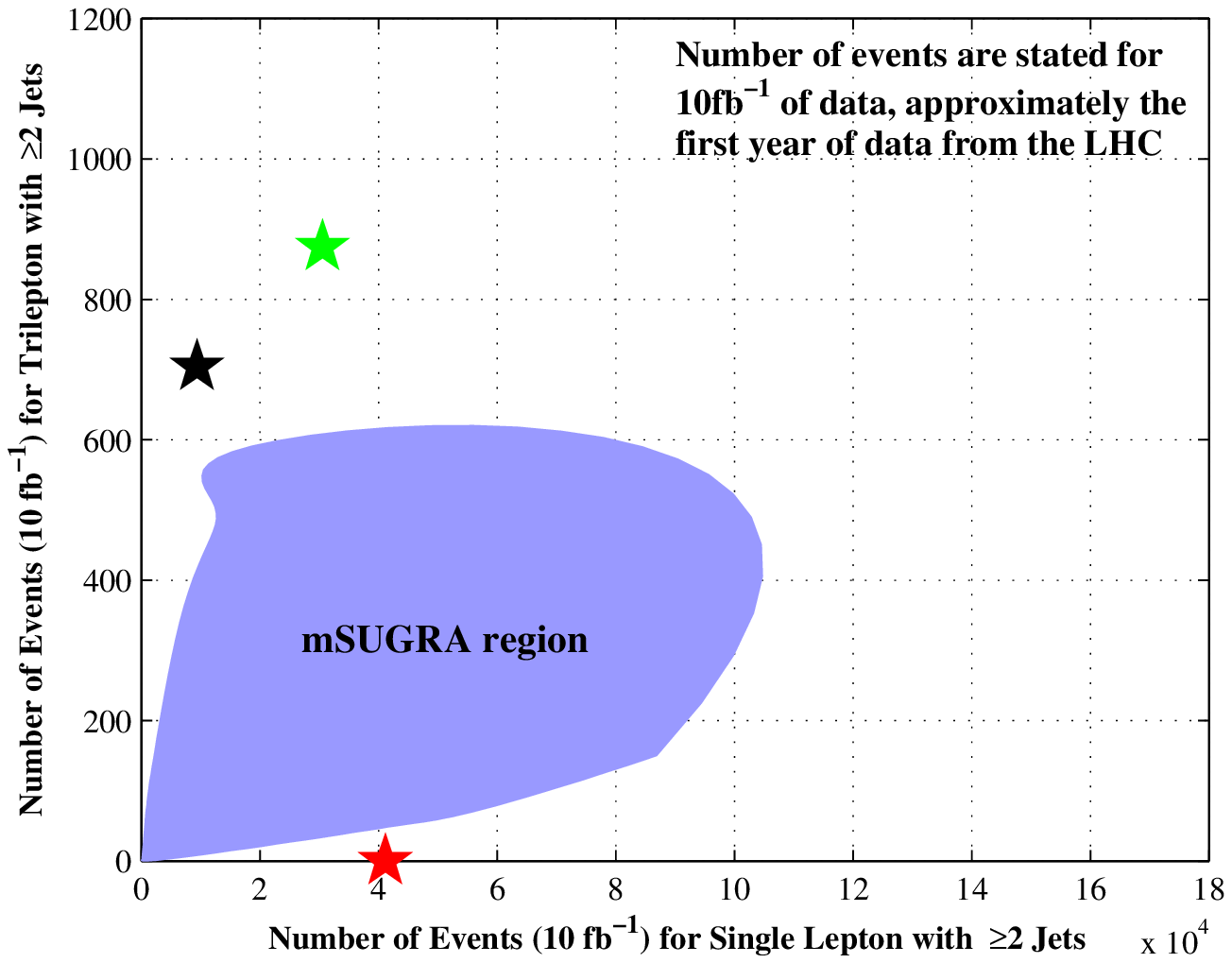,height=12cm, angle=0}
\caption{Number of events for Trilepton (3L) with $\geq$ 2
jets vs Number of events for Single Lepton (1L) with $\geq$ 2 jets. The
blue area denotes the region occupied by mSUGRA, while the red, green
and black ``stars'' correspond to other supersymmetry breaking models.}
\label{msugra:fig:OSvs0L} 
\end{figure}

\begin{figure}
\center \epsfig{file=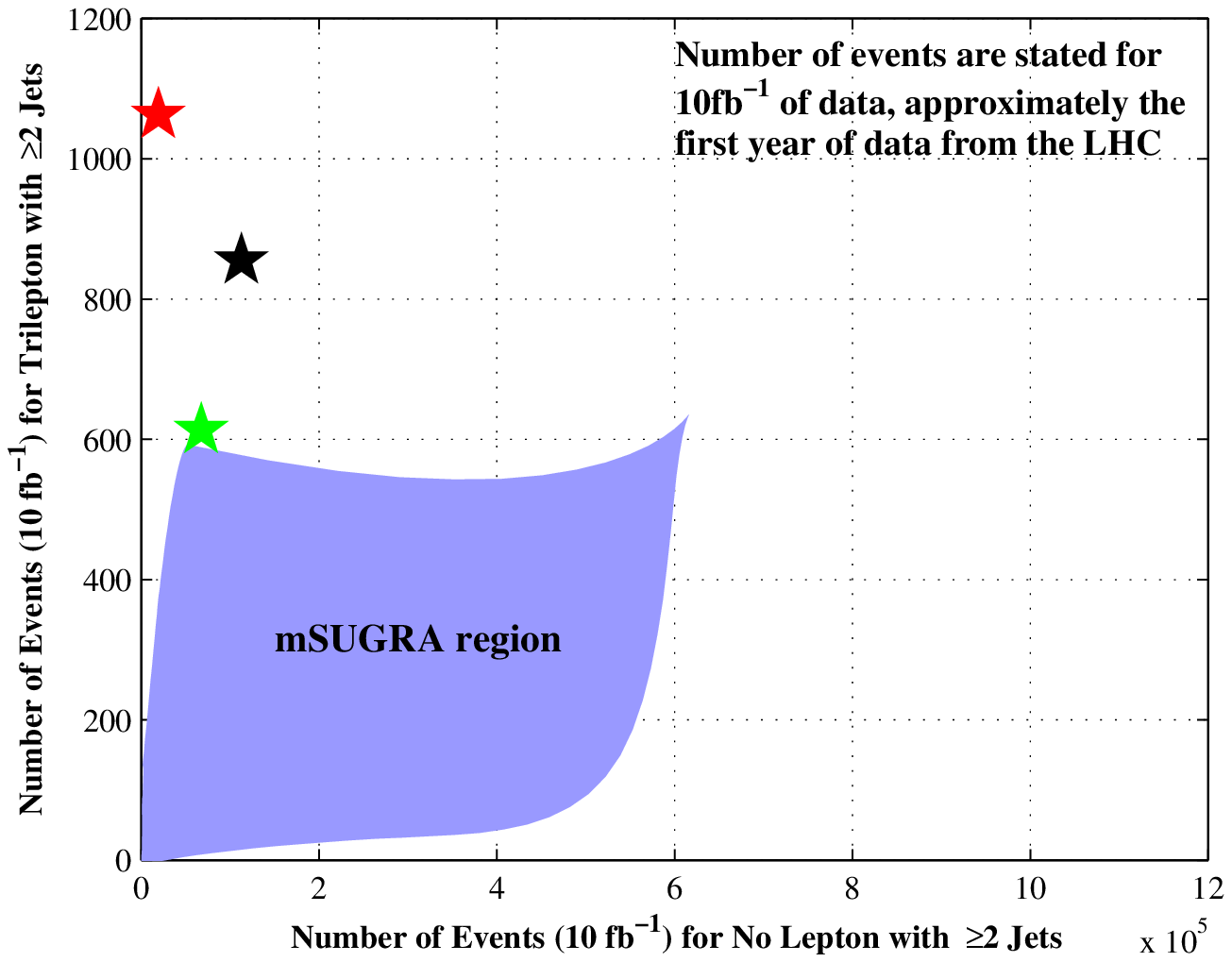,height=12cm, angle=0}
\caption{Number of events for Trilepton (3L) with $\geq$ 2
jets vs Number of events for No lepton (0L) with $\geq$ 2 jets. The blue area denotes the 
region occupied by mSUGRA, while the red, green and black ``stars''
correspond to other supersymmetry breaking  
models.} \label{msugra:fig:OSvs0L}
\end{figure}

The blue area in these plots denotes the experimentally allowed region
occupied by mSUGRA for any possible parameters. The ``stars'' denote other models of supersymmetry breaking and are present
only to illustrate that it is easy to be `outside the mSUGRA box'. 
It should be kept in mind that while theory predicts the cross-section for a given process, experiments only measure the number of events \footnote{\noindent An integrated luminosity of 10 $fb^{-1}$ has been assumed for the analysis; this corresponds roughly to a luminosity of $10^{33}\,\mathrm{cm}^{-2}\,s^{-1}$ for one collider year ($\sim 10^{7}$ s).}.  Therefore, to compare explicit theoretical predictions with experimental measurements, the luminosity for the process has to be known precisely. This is quite hard to do in practice and luminosity determination is accompanied by large uncertainties. Our approach is useful even if the luminosity is not known. One way to implement this is to
normalize the number of events for all signatures with respect to a reference signature, thus plotting ratios of the number of events with respect to that of the reference signature, for both experiment and theory. This will help in reducing the dependence of the inclusive signatures on the luminosity. This has not been done in our analysis for simplicity, but should be done in more sophisticated analyses.

From the plots, it is clear that the mSUGRA region occupies a
\emph{bounded} region in inclusive signature space. This confirms our
expectation, as was already mentioned in section 4. Since the mSUGRA
parameter space is a smooth region, it is reasonable to expect that it would
map to a smooth region in inclusive signature space. Therefore,
even though in practice a set of about ten thousand ``discrete'' points was used for the 
simulation, the mSUGRA regions have been represented as a smooth
continuous ``blobs''. In general, there is statistical error since we have sampled a finite
number of mSUGRA models. In addition, even though we have imposed
a minimal set of cuts, the shape and size of the regions may
depend somewhat on the cuts imposed. The blobs have been drawn such that they
\emph{overestimate} the regions occupied by mSUGRA, to make results more robust. Overestimating the set of
points as smooth blobs helps us to get a rough idea of the ``size''
of the mSUGRA region in each plot. For real data with experimental cuts, the relevant region can be
determined as precisely as desired.

We should emphasize again that our goal is to illustrate the advocated approach by a simple
example. Of course, more sophisticated analysis can (and should)
be done, once we get a better understanding of the basics. For
example, more sophisticated cuts which take into account the
actual detector environment can be imposed. However, even a
simplified analysis can teach us many things. First, in each of the plots,
three points are shown which lie ``far away'' from mSUGRA. These points
correspond to other models of supersymmetry breaking. The precise
nature of these models is \emph{unimportant} for this analysis. They have been shown here
merely as ``existence proofs'' of models which occupy a different region in
inclusive signature space. It is undoubtedly important to understand the reasons why 
other supersymmetry breaking approaches (models), such as those represented by the ``stars'' in 
the plots, lie outside the mSUGRA region. However, that is not the emphasis of the present paper 
and will be analyzed in future work.

Second, as has been mentioned before, in the event of a discovery
at the LHC, this method could rule out mSUGRA. Signals observed
will appear as isolated points in each of the plots. If even a
single point lies outside the mSUGRA region in any of the plots,
then mSUGRA has to be ruled out. This is quite clear. However, the
situation is not so simple if \emph{all} the points lie within the
mSUGRA regions. It does not imply that mSUGRA has been confirmed.
This is because, in general, there could be an overlap in the
regions occupied by different models \footnote{\noindent We have explicitly
seen this in our analysis of other models.}. In such a situation, further
analysis needs to be done. One possibility is to better define
the concept of ``distance'' --- to a particular model from an
experimental data point or between two models, in inclusive
signature space. It would then seem reasonable to postulate that
models which are closer in ``distance'' to the experimental data
point than others in inclusive signature space, are more favorable
than others. However, sufficient care has to be taken in defining
such a concept so that subtle effects involving different types of
uncertainties and errors are properly taken into account. A preliminary study 
of these kinds of issues has been done in \cite{Allanach:2004my}.

\section{Generalization}
It is clear that our approach can shed important light on the status
of mSUGRA in the event of a discovery at the LHC. However, we wish to go beyond mSUGRA and
generalize our approach in ways such that it is possible to reap its maximum benefits.

More generally, the approach can be applied to study many questions and implications of any
hadron collider signal. It can also be used to systematically combine collider and
non-collider information. A natural thing would be to do the same analysis for a variety of
other supersymmetry breaking models. That is, one should try to
carve out the parameter space of different supersymmetry breaking
models and study their inclusive signatures. This in turn has the
potential to distinguish and ultimately rule out entire classes
of models. Different models will tend to occupy different 
bounded regions in the space of inclusive signatures. Even though
the regions occupied by some or many of these models might overlap
in some or many of the plots, they are bound to be different in at
least some of them. Detected signals will appear as isolated
points in the relevant inclusive signature plots. The models whose
``regions'' in \emph{all} the relevant inclusive signature plots
contain the point, are compatible with the experimental data while
other models are not. From the argument of the previous section,
it requires more work to differentiate models which are
``compatible'' with the experimental data in the above sense.
Another way to differentiate models would be to introduce new
inclusive signatures (or combinations of inclusive signatures from
different types of experiments) which provide new information. 

One could go even further. Our theoretical prejudice orients us
towards supersymmetry breaking models which explain supersymmetry
breaking from a fundamental theory (like string theory), because
they have an ultraviolet completion in contrast to others. Various
aspects of low energy phenomenology could crucially depend on
specific string constructions (for example, the heterotic vs D-brane
constructions), on specific moduli fields acquiring vacuum
expectation values (the dilaton, kahler moduli or complex
structure moduli) and on the values of numbers and charges that
appear in string theory (modular weights, anomalous U(1) charges,
anomaly cancellation coefficients, topological quantities, etc.)
to name a few. Armed with our approach, trying to figure out
general ``rules of thumb'' about how the above alternatives affect
inclusive signatures would be of great value in choosing the most
promising avenue for further top-down study.

In addition, from a more bottom-up perspective, one could try to study the entire parameter
space of the MSSM and its effect on the pattern of inclusive signatures. Of course, a brute force
analysis is a herculean computation challenge. However, once we understand
different corners of the parameter space of the MSSM (mSUGRA for example) and their effect on the
pattern of inclusive signatures, an intelligent intuitive approach can allow us to make important statements
about the effects of a generic MSSM model on the pattern of inclusive signatures.

\section{Conclusion}
In this paper, we have emphasized some obstacles to
determining the implications of new physics at the LHC. Proper
study of the obstacles in a general theory is
lacking in the literature, as most phenomenological analyses rely
on explicit and implicit assumptions. In light of these obstacles and the
incomplete nature of theoretical models at our disposal, an
innovative and constructive approach is required. One such
approach has been presented in the paper - to learn about the
properties of new physics from \emph{patterns} of inclusive
signatures.

This approach has been applied here to the case of the
minimal supergravity mediated supersymmetry breaking scenario
(mSUGRA) in detail, resulting in a ``footprint'' of mSUGRA in
``inclusive signature space'', seen from figures 1-4 in the paper.
The figures clearly show that mSUGRA occupies a finite bounded region
in inclusive signature space and it is possible to find models which lie outside
this region. Thus, in the event of a discovery, this
approach can rule out mSUGRA if any experimental point lies outside the mSUGRA
region in any inclusive signature space projection. However, confirming
mSUGRA (or any particular model) requires further
analysis.

The signature space approach also has major experimental advantages, allowing reduction
of the effects of luminosity uncertainties, and of acceptance and jet energy corrections.

It is possible to generalize the approach to other models, distinguish
between them and to get valuable insights about different approaches to
supersymmetry breaking and the fundamental microscopic theory which leads to it.
The main purpose of this paper was to illustrate the
above approach by applying it to a particular model --- mSUGRA,
the most popular ``model''  for phenomenological studies. A
more sophisticated analysis can (and should) be done when data is available, taking
various effects like detector environment, errors, statistics,
etc. into account.

\vspace*{1cm}
\noindent{\Large \bf Acknowledgement} \\
The authors appreciate helpful conversations with and suggestions
from Lian-Tao Wang and Brent Nelson. GLK also appreciates discussions with Nima Arkani-Hamed. 
The research of JLB, GLK, PK and TTW is
supported in part by the U.S. Department of Energy.

\end{document}